# Impact of edge turbulence spreading on broadening the heat flux width with plasma approaching the density limit

T. Wu[1], P.H. Diamond[2, *], L. Nie[1], R. Ke[1], Z. P. Chen[3], Q. H. Yang[3], W. J. Tian[1], T. Long[1], Z. J. Yang[3], Z.Y. Chen[3], M. Xu[4, *]

[1] Southwestern Institute of Physics, Chengdu 610225, China

[2] Departments of Astronomy, Astrophysics and Physics, University of California, San Diego, CA 92093, United States of America

[3] State Key Laboratory of Advanced Electromagnetic Technology, International Joint Research Laboratory of Magnetic Confinement Fusion and Plasma Physics, School of Electrical and Electronic Engineering, Huazhong University of Science and Technology, Wuhan 430074, China

[4] Institute of Modern Physics, Fudan University, Shanghai 200433, China

E-mail: pdiamond@ucsd.edu and minxu_mx@fudan.edu.cn

## Abstract

This paper investigates the impact of edge turbulence spreading on broadening the heat flux width in Ohmic-plasma approaching the density limit of the J-TEXT tokamak. At the plasma edge, the $\mathbf{E} \times \mathbf{B}$ shear flow collapses while turbulence transport and spreading enhances significantly when approaching the density limit. The heat flux width increases with normalized density. An energy production ratio model is used to quantify the contribution of edge turbulence spreading to the origin of the SOL turbulence. Experimental data show that the energy production ratio is much larger than 1, indicating that turbulence spreading at separatrix is the origin of the SOL turbulence. The heat flux widths increase with edge turbulence spreading as well as the energy production ratio. The impact of blob-induced transport on the heat flux width is investigated in detail. Especially, the average blob-induced spreading is about 81% of the total edge spreading in the high-density scenario. Blobs with larger radial scales enhance edge spreading into the SOL, thus dominating the SOL turbulence and consequently broadening the heat flux width. These results suggest that edge turbulence spreading plays a crucial role in broadening the heat flux width as plasma approaches the density limit.

# 1 Introduction

The distribution of the heat load on the divertor targets is a critical issue for International Thermonuclear Experimental Reactor (ITER) and future magnetic fusion reactors. Concerns about this issue drive measurements and predictions of the heat flux width $\lambda_q$ [1] in various plasma scenarios. The magnitude and scaling trends of $\lambda_q$ are of great interest [2–18]. For the ITER start-up phase, scaling laws of $\lambda_q$ in limited Ohmic-plasma have been studied [6–10, 15].

High density scenarios are expected for fusion reactors, as plasma density is an important element of fusion triple product and fusion power is proportional to the square of plasma density. In the high-density scenario, turbulence and turbulent transport increase at the plasma edge [19–22]. Note, 'edge' in this paper refers to the radial position just inside the last closed flux surface (LCFS). When plasma density approaches the density limit, edge cooling, decreased $\mathbf{E} \times \mathbf{B}$ shearing rate and enhanced turbulence transport at the plasma edge and confinement degradation may happen [23–25]. On the J-TEXT tokamak edge $\mathbf{E} \times \mathbf{B}$ shear flow collapses and turbulent particle transport increases [26–29]; edge turbulence spreading [30–34], as well as blob transport [35–36] increase significantly [26, 28] when approaching the density limit. The heat flux width is reported to be broadened by increased edge turbulence in the high-density scenario. [37–40]. The role of collisionality is emphasized to enhance edge turbulence and hence to broaden the heat flux width [37–40]. However, the enhanced edge turbulent transport in the high-density scenario and its role in the SOL turbulence have not been studied in plasma density approaching the density limit. In addition, increased blob transport has been observed in the high-density scenario, but the impact of blob transport on broadening of the heat flux width has rarely been investigated. This paper aims to study the impact of the edge turbulent transport and blob transport on broadening the heat flux width with plasma approaching the density limit.

Essentially, the heat flux width is determined by the balance between parallel and perpendicular transport in the scrape-off layer (SOL) [1]. It is already known that edge turbulent transport, especially the edge turbulence spreading plays a significant role in broadening the heat flux width [41–44]. However, the role of edge turbulence spreading on the SOL perpendicular transport requires further study. The energy production ratio model [43], has clarified the contribution of edge turbulence spreading across the LCFS to the origin of SOL turbulence. This model assumes that SOL turbulence originates mainly from edge turbulence spreading and SOL local interchange turbulence production. The edge turbulence spreading is deduced from an effective gravity by field fine curvature in the equation of motion. The energy production ratio model is defined as:

$$R_a \equiv \frac{\Gamma_{E,LCFS}}{\int_0^{\lambda_q}\left[\frac{C_s^2}{R}\langle\widetilde{V}_r\frac{\widetilde{n}_e}{n_e}\rangle - \langle\widetilde{V}_\theta\widetilde{V}_r\rangle\frac{\partial\langle V_\theta\rangle}{\partial r}\right]dr} \quad (1)$$

Here, $\Gamma_{E,LCFS}$ is the turbulence spreading, i.e. the flux of turbulence energy across the LCFS. The denominator in equation 1 is an integral (from the LCFS to $\lambda_q$) of the total SOL production of turbulence. It consists of a local production term from the interchange instability and a damping term due to Reynolds power. If the Reynolds power is much smaller than the local interchange term, then $R_a$ can be simplified as:

$$R_a \cong \frac{\Gamma_{E,LCFS}}{\int_0^{\lambda_q}\left[\frac{C_s^2}{R}\langle\widetilde{V}_r\frac{\widetilde{n}_e}{n_e}\rangle\right]dr} \quad (2)$$

In reference [43], $R_a < 1$ where the edge geodesic acoustic mode provides a relatively strong $\boldsymbol{E_r \times B}$ shearing rate, hence local SOL interchange turbulence dominates the SOL turbulence. With a relatively weak $\boldsymbol{E_r \times B}$ shearing rate in the edge, $R_a > 1$ and the heat flux width increases with $R_a$, indicating the dominant contribution of edge turbulence spreading to the origin of SOL turbulence, and the significant role of edge turbulence spreading in broadening the heat flux width.

In this paper, we use the energy production ratio model to investigate the impact of edge turbulence spreading on broadening the heat flux width with plasma approaching the density limit. The experiments were conducted on J-TEXT limited Ohmic-plasma. The increasing plasma density approaching the density limit provides a natural range of variation of the edge $E_r \times B$ shear flow. Regarding the density limit, the edge $E_r \times B$ shear flow collapses, and meanwhile the edge turbulent transport and turbulence spreading enhance substantially. The experimental heat flux widths are several times of the predicted widths by the generalized Heuristic Drift (GHD) model [38]. The energy production ratio $R_a$ is much larger than 1, indicating that edge turbulence spreading is the origin of SOL turbulence. Furthermore, the impact of blob-induced transport on broadening the heat flux width with increasing density is investigated systematically. Blob-induced spreading dominates the total edge spreading and hence the SOL turbulence, thus broadening the heat flux width. These results suggest that edge turbulence spreading plays a significant role in broadening the heat flux width when plasma approaches the density limit.

The remainder of this paper is organized as follows. Section 2 describes the experimental arrangement, including the database and diagnostics. Section 3 reports the plasma profiles and basic fluctuation properties with increasing density. Section 4 presents the heat flux width broadening by edge transport and comparison with widths from GHD model. The impact of edge turbulence spreading on the heat flux width is investigated in section 5. Edge turbulence spreading by blobs and their roles in the heat flux widths are elucidated in section 6. Section 7 presents conclusions and further study.

## 2     Experimental Arrangement

*2.1 Plasma discharge parameters and database*

The experiments were conducted in Ohmically-heated hydrogen plasmas on the J-TEXT tokamak [45–46]. J-TEXT has major and minor radii of $R = 1.05$ m and $a = 0.22$ m, respectively. The cross-section is circular, with a limiter configuration shown in Figure 1. There are 3 poloidal limiters in at port 14 on J-TEXT, which is similar to a single poloidal ring limiter, as shown in Figure1b. The parallel connection length is estimated as $L_\parallel \approx \pi R = 3.3$ m [1, 47]. The effects of different limiter configurations on the heat flux width are studied in the references [48-49]. The discharge numbers are 1094988-1095026 in the same day. The main plasma parameters are: plasma current $I_p = 105$ kA, toroidal magnetic field $B_t = 2.1$ T, loop voltage $V_L = 2.1$ V, safety factor $q_a = 4.2$, line-average density $\bar{n}_e = 2.5 - 4.1 \times 10^{19}$ m$^{-3}$. The Greenwald density limit is calculated $n_G =$ $I_\mathrm{p}/\pi a^2 = 6.9 \times 10^{19}$ m$^{-3}$. The ratio of the line-average density to the Greenwald density limit, which is referred to as the normalized density $\bar{n}_\mathrm{e}/n_\mathrm{G} = 0.36 - 0.60$. Gas puff is used to increase plasma density in these experiments.

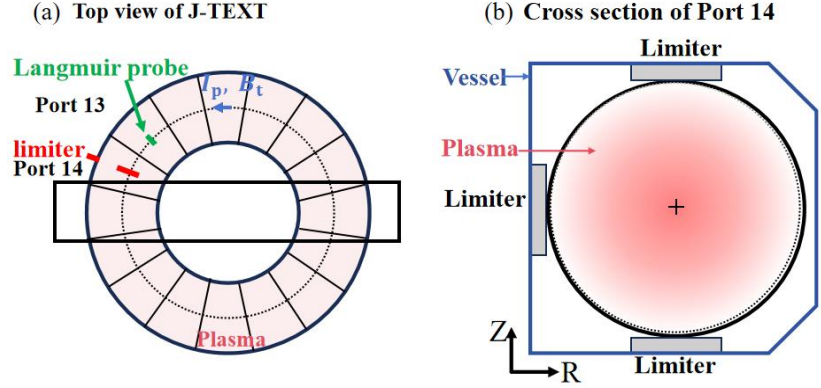

Figure 1. (a) Schematic top view of J-TEXT tokamak and (b) Cross-section of J-TEXT tokamak with locations of the 3 limiters at port 14.

The database includes 32 shots in total. These data are divided into 3 sets according to their different density ratios, i.e., $\bar{n}_\mathrm{e}/n_\mathrm{G}=0.36-0.45$, $\bar{n}_\mathrm{e}/n_\mathrm{G}=0.45-0.52$ and $\bar{n}_\mathrm{e}/n_\mathrm{G}=0.52-0.60$, respectively. Relevant information is listed in Table 1. The heat flux widths in these shots are in the range of 8 – 45 mm. The mean $\lambda_\mathrm{q}$ values are $13 \pm 1.4$ mm, $20 \pm 2.0$ mm, and $34 \pm 1.6$ mm for the 3 different density ratios. The electron collision frequency at the LCFS $\nu_\mathrm{e} = 1/(3(2\pi)^{3/2} \frac{\varepsilon_0^2 m_\mathrm{e}^{1/2} T_\mathrm{e}^{3/2}}{n_\mathrm{i} Z^2 e^4 \ln\Lambda}) \approx 1 - 8 \times 10^6 \mathrm{s}^{-1}$. The effective collisionality is $\nu_{\mathrm{e}*} = 6.921 \times 10^{-18} \frac{qRn_\mathrm{e}Z\ln\Lambda_\mathrm{e}}{T_\mathrm{e}^2 \epsilon^{1.5}} \approx 20 - 300$ at the LCFS.

Table 1. The details of 3 sets in database.

| Classification | $\bar{n}_\mathrm{e}$ ($10^{19}$m$^{-3}$) | $\bar{n}_\mathrm{e}/n_\mathrm{G}$ | Number of shots | Exp $\lambda_\mathrm{q}$ (mm) | Mean $\lambda_\mathrm{q}$ (mm) |
|---|---|---|---|---|---|
| Low-density scenario | 2.5 – 3.1 | 0.36 – 0.45 | 5 | 8 – 16 | $13 \pm 1.4$ |
| Medium-density scenario | 3.1 – 3.6 | 0.45 – 0.52 | 14 | 12 – 36 | $20 \pm 2.0$ |
| High-density scenario | 3.6 – 4.1 | 0.52 – 0.60 | 13 | 26 – 45 | $34 \pm 1.6$ |

*2.2 Diagnostic setup*

The fast-reciprocating Langmuir probe arrays are used in these experiments. The configuration of the Langmuir probe is a 4-tip probe [50]. The sampling rate is 2 MHz. The fluctuation of poloidal electric field $\widetilde{E}_\theta$ is estimated by the poloidal floating potential difference, when neglecting the effects of electron temperature fluctuations. Radial velocity fluctuation is estimated as $\widetilde{V}_r \approx \widetilde{E}_\theta/B_t$ [43]. The turbulent parameters are calculated as: turbulent particle flux is $\Gamma_r = \langle \widetilde{V}_r \widetilde{n}_e \rangle$, the radial flux of turbulence internal energy $C_s^2 \langle (\widetilde{n}_e/n_e)^2 \widetilde{V}_r \rangle$, with $C_s$ the ion sound velocity [43] and the turbulence spreading rate $\omega_s = -\frac{1}{2}\frac{\partial \langle \widetilde{n}_e^2 \widetilde{V}_r \rangle}{\partial r} / \frac{1}{2}\langle \widetilde{n}_e^2 \rangle$ [33]. Plasma potential is inferred by the expression $V_p = V_f + \alpha T_e$, with $\alpha = 2.4$ as the sheath coefficient [26]. The radial electric field is calculated by $E_r = -\nabla_r V_p$, and the shearing rate of the mean $\mathbf{E_r} \times \mathbf{B}$ flow is $\omega_{E_r \times B} = |\nabla_r \left(\frac{E_r \times B_T}{B_T^2}\right)|$. The parallel heat flux can be expressed as $q_\parallel = \gamma J_s T_e$ (here $J_s = I_s/A_{\text{eff}}$) with $\gamma$ electron sheath heat transmission coefficient ($\gamma = 7$ for T-TEXT Ohmic-plasma with a limiter configuration [1]). The radial profile of $q_\parallel$ is assumed to follow a single exponential decay in the SOL, i.e., $q_\parallel(r) = q_{\parallel,\text{LCFS}}\exp(-r/\lambda_q)$, where $q_{\parallel,\text{LCFS}}$ is the parallel heat flux at the LCFS, and $r$ is the distance from the LCFS. Hence $\lambda_q$ is estimated as $\lambda_q = -\left(\frac{\partial \ln(q_\parallel(r))}{\partial r}\right)^{-1}$ via the log-linear fit method [41, 43]. The electron adiabaticity parameter is calculated as $\alpha = k_\parallel^2 V_{\text{th},e}^2/\nu_{ei}\omega$, where $k_\parallel$ is the parallel wavenumber ($k_\parallel \approx 1/Rq$), $V_{\text{th},e}$ is the electron thermal velocity, $\nu_{ei}$ is the electron-ion collision rate, and $\omega$ is the dominant turbulence frequency [26].

## 3   Plasma profiles and turbulence properties with increasing density

*3.1 Plasma profiles in the boundary*

The radial profiles of plasma parameters in the boundary are examined in this section. Radial profiles are average values from 3 data sets (Table 1) in $-10 \text{ mm} \leq r - r_{\text{LCFS}} \leq 30 \text{ mm}$ in Figure 2. With increasing plasma density, the electron density changes slightly while the electron temperature and parallel heat flux decrease in the edge and near SOL. Radial gradients of electron density, electron temperature, and parallel heat flux decrease in the edge and near SOL. The electron density and electron temperature profiles in the whole cross section can be refereed to [26]. The edge poloidal $\mathbf{E_r} \times \mathbf{B}$ velocity $V_{E_r \times B}$ and its shear decrease with increasing plasma density. Especially, in the high-density scenario approaching the density limit ($\bar{n}_e/n_G \approx 0.52 - 0.60$), the $\mathbf{E_r} \times \mathbf{B}$ shear flow collapses, and the relative fluctuation $\tilde{I}_s^{rms}/I_s$ is much larger than that with lower density at the plasma edge. In the SOL, the turbulent particle flux decreases in the high-density scenario while it increases and then decreases in the low and medium density scenarios in Figure 2(f). Note that the error bars in figures 2 present a confidence interval of 68%. These results are consistent with collapsed $E_r \times B$ shear flow and enhanced turbulent transport at plasma edge when plasma approaches density limit on J-TEXT [26-29].

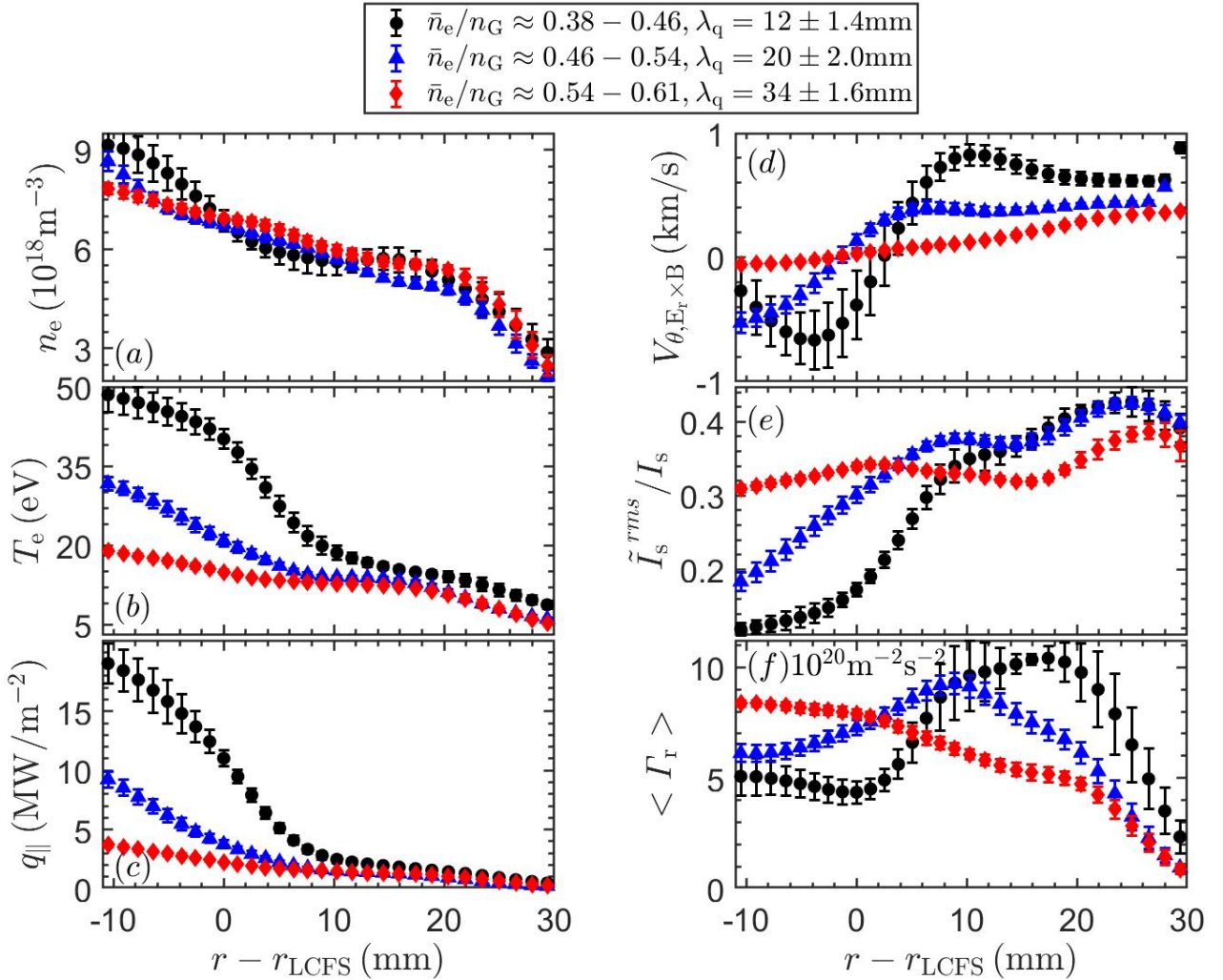

Figure 2. The radial profiles of (a) electron density $n_e$, (b) electron temperature $T_e$, (c) parallel heat flux $q_\parallel$, (d) $E_r \times B$ velocity $V_{E_r \times B}$, (e) relative fluctuation of ion saturation current $\tilde{I}_s^{rms}/I_s$ and (f) turbulent particle flux $<\Gamma_r>$ in the plasma boundary with different normalized density $\bar{n}_e/n_G$.

*3.2 Turbulent parameters with increasing density*

We examine the turbulent parameters with increasing plasma density approaching the density limit on J-TEXT. Parameters in Figures 3-4 are measured near the LCFS. Figure 3(a) shows that the $E_r \times B$ shearing rate decreases with increasing $\bar{n}_e/n_G$. Figure 3(b) shows the turbulence decorrelation rate $\omega_{de}$ (the reciprocal of turbulence auto-correlation time [28]) decreases with $\bar{n}_e/n_G$, indicating that the turbulence auto-correlation time increases with $\bar{n}_e/n_G$. The ratio of the $\mathbf{E_r \times B}$ shearing rate to the turbulence decorrelation rate $\omega_{E_r \times B}/\omega_{de}$ decreases with $\bar{n}_e/n_G$ in Figure 3(c). Moreover, $\omega_{E_r \times B}/\omega_{de} \ll 1$ in the high-density scenario, indicating that edge $\mathbf{E_r \times B}$ shear flow collapses and cannot suppress the edge turbulence and turbulent transport. The reduction in figures 3(a) and 3(b) indicates that it is the edge $\mathbf{E_r \times B}$ shear flow, rather than normalized $\omega_{E_r \times B}/\omega_{de}$, plays important role in the edge turbulence. The relative fluctuation, turbulent particle flux and radial flux of turbulence internal energy near LCFS are increasing with $\bar{n}_e/n_G$. In addition, edge turbulence is in the electron diamagnetic drift direction for all these discharges.

The electron adiabaticity is investigated with increasing normalized density in Figure 4. Plasma converts from the adiabatic regime ( $\alpha > 1$ ) to hydrodynamic regime ( $\alpha < 1$ ) with increasing normalized density. This is consistent with the results that plasma changes to the hydrodynamic regime when plasma approaches the density limit [26–28].

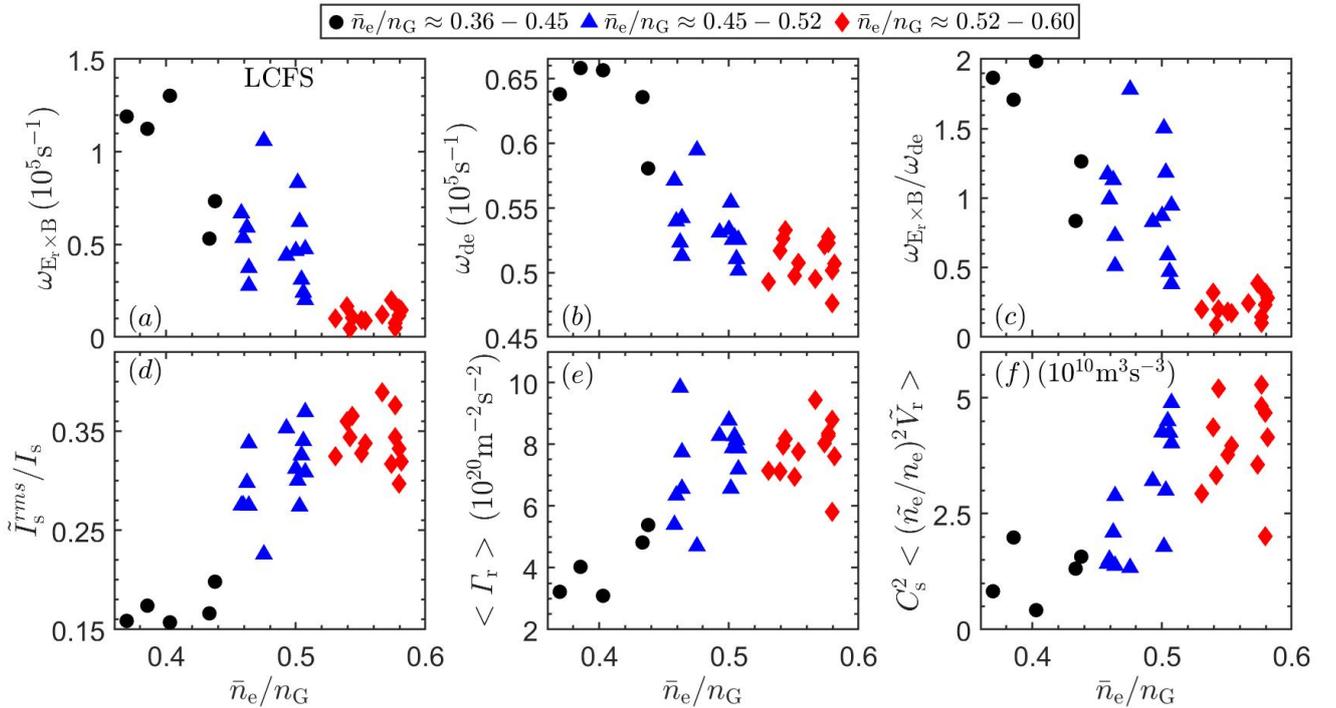

Figure 3. (a) The $\mathbf{E_r \times B}$ shearing rate $\omega_{E_r \times B}$, (b) the turbulence decorrelation rate $\omega_{de}$ and (c) the ratio of the $\mathbf{E_r \times B}$ shearing rate to the turbulence decorrelation rate $\omega_{E_r \times B}/\omega_{de}$, (d) the relative fluctuation $\tilde{I}^{rms}_s/I_s$, (e) turbulent particle flux $\langle \Gamma_r \rangle$ and (f) the radial flux of turbulence internal energy near LCFS with increasing normalized density

$\bar{n}_\mathrm{e}/n_\mathrm{G}$.

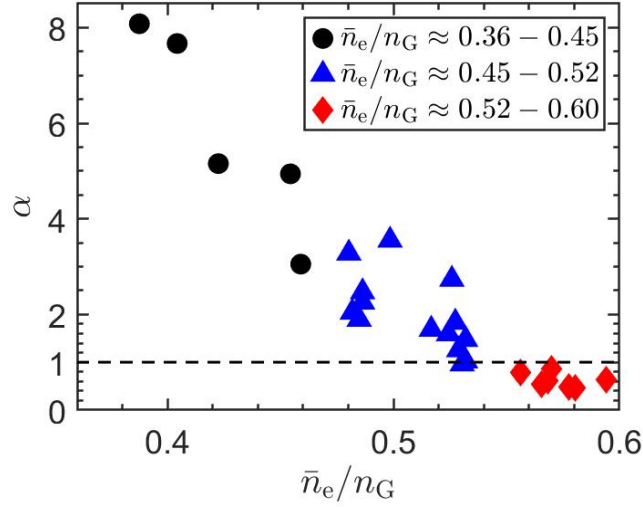

Figure 4. The electron adiabaticity $\alpha$ with increasing normalized density $\bar{n}_\mathrm{e}/n_\mathrm{G}$.

## 4    The heat flux widths broadening

*4.1 Correlations between the heat flux widths and turbulent parameters*

The correlations between $\lambda_\mathrm{q}$ with turbulent parameters are investigated in Figures 5 and 6. $\lambda_\mathrm{q}$ is negatively correlated with the $\boldsymbol{E_\mathrm{r}} \times \boldsymbol{B}$ shearing rate in Figure 5(a), consistent with [41, 43, 51–52]. $\lambda_\mathrm{q}$ correlates positively with relative fluctuation $\tilde{I}_\mathrm{s}^{rms}/I_\mathrm{s}$, turbulence auto-correlation time and turbulent particle flux. This agrees with previous results in [41, 43]. These results agree with previous findings that heat flux widths increase with edge turbulence and turbulent transport while decreases with the $\boldsymbol{E_\mathrm{r}} \times \boldsymbol{B}$ shearing rate [41–44, 51–58].

The correlations of heat flux widths with the effective collisionality near the LCFS and SOL collisionality are shown in Figure 6. Heat flux widths increase with the effective collisionality near the LCFS, consistent with results in [39]. The SOL collisionality is defined as $v^*_\mathrm{SOL} = 10^{-16} n_\mathrm{u} L_\parallel / T_\mathrm{u}^2$, where $n_\mathrm{u}$ is the separatrix density and $T_\mathrm{u}$ is the temperature in the upstream. The SOL collisionality is less than 15, which confirms that plasma is in the sheath-limited regime with small temperature gradient [1]. The SOL collisionality at the LCFS increases with normalized density as well as the SOL collisionality.

Under almost the same discharge condition except increasing line-average density, the heat flux widths in these experiments are in the range of 8–45 mm. In the high-density scenario approaching the density limit, the average heat flux width is 2.5 times larger than that in the low-density scenario. These experimental results manifest that enhanced edge turbulence and turbulent transport in plasma approaching the density limit can broaden the heat flux width significantly. We will study further the broadening of the heat flux width in plasma approaching the density limit by comparing experimental results with the prediction from GHD model in the next section.

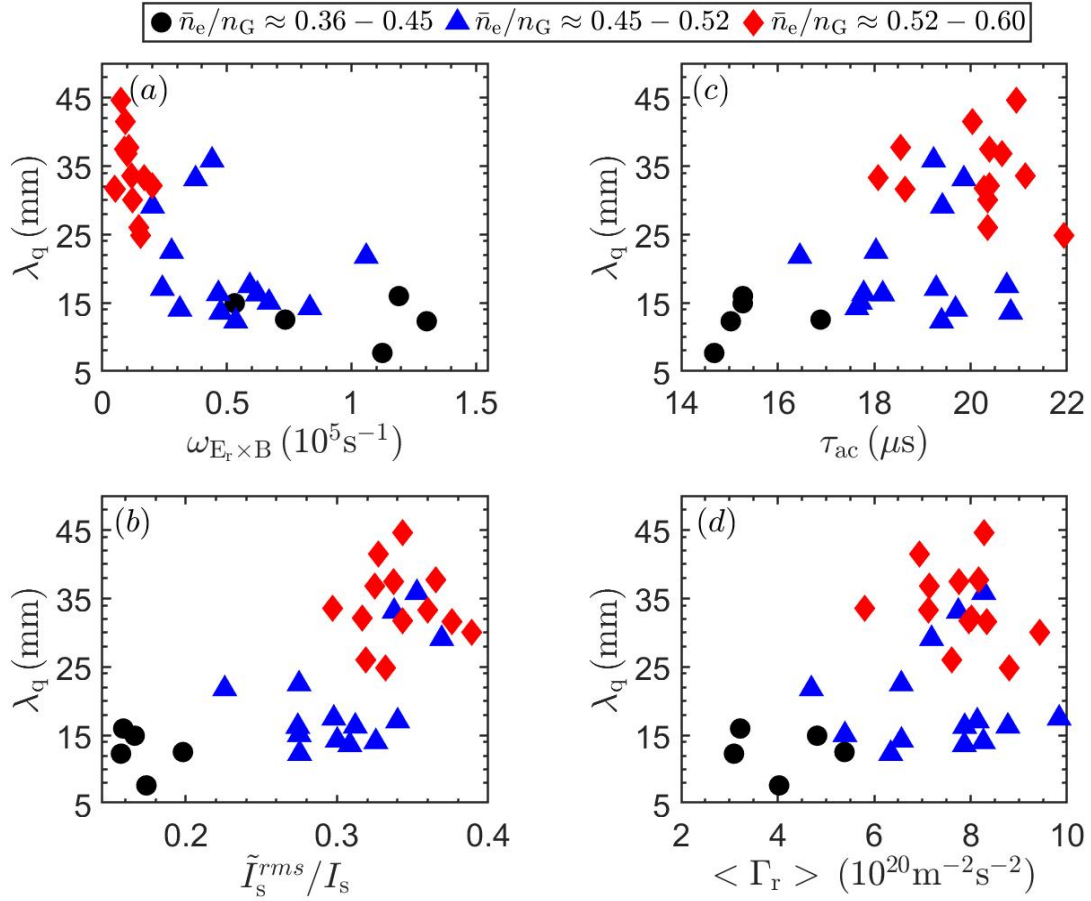

Figure 5. The correlations between heat flux widths $\lambda_q$ and (a) the $\boldsymbol{E_r \times B}$ shearing rate, (b) relative fluctuation $\tilde{I}_s^{rms}/I_s$, (c) turbulent particle flux and (d) turbulent particle flux diffusion coefficient with increasing normalized density $\bar{n}_e/n_G$.

.

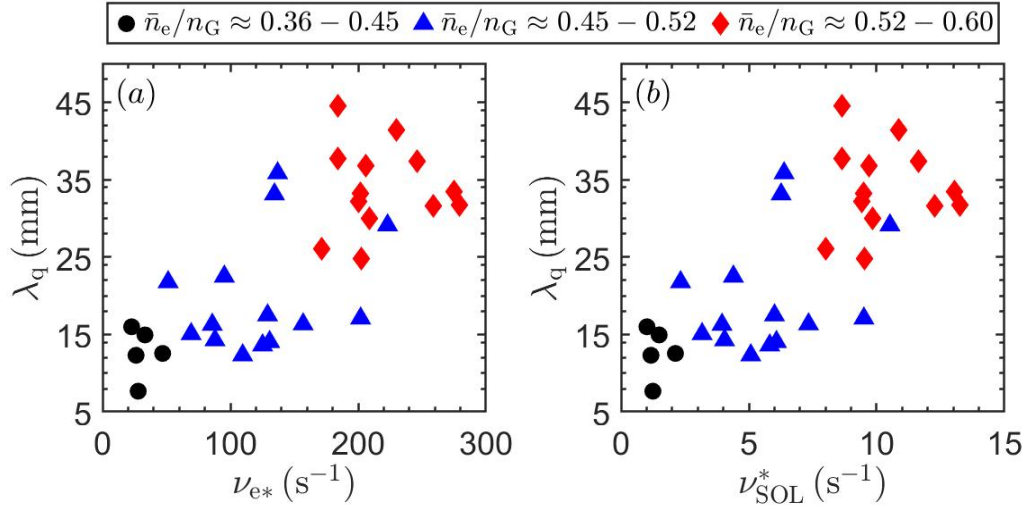

Figure 6. The correlations between the heat flux widths $\lambda_q$ and (a) effective collisionality $\nu_{e*}$ near the LCFS and (b) SOL collisionality $\nu^*_{SOL}$ measured at the LCFS with increasing normalized density $\bar{n}_e/n_G$.

*4.2 Heat flux widths from experiments and the GHD model*

The heat flux widths from experiments and the generalized heuristic drift (GHD) model [38] are compared. The average heat flux width predicted by the heuristic drift (HD) model [4] is $\lambda_{q,HD} \approx 2\epsilon\rho_{\theta,i} = 3$ mm. The HD model is based on neoclassical drift assuming no turbulence in the SOL region. The GHD model takes the finite SOL collisionality into account [38]. The SOL residence time is estimated by the parallel energy confinement time instead of the parallel particle confinement time. Thus, the GHD model predicts a collisional broadening factor over the HD value $\lambda'_q$ from $\lambda'_q\left(1-(\frac{1-f_{power}}{\lambda'_q})^7\right) = 0.0739\eta\nu^*_{SOL}$, where $\eta$ is the order-unity factor $\eta = \left(\frac{1+Z}{A}\right)^{\frac{1}{2}}f(Z_{eff})$, $f(Z_{eff}) = 0.672 + 0.076Z_{eff}^{1/2} + 0.252Z_{eff}$ is a fit to the Braginskii electron thermal conductivity as a function of effective charge number $Z_{eff}$, and $f_{power}$ is the fraction of power entering the SOL that is lost to radiative and atomic process near the divertor target. In these experiments, heat flux width is measured at the mid-plane and $f_{power} = 0$.

The relation of $\lambda'_q$ to the normalized density and the ratios of experimental heat flux widths to HD width $\lambda_q/\lambda_{q,HD}$ versus $\lambda'_q$ are elucidated in Figure 7. $\lambda'_q$ and $\lambda_q/\lambda_{q,HD}$ increases with plasma density, consistent with results in [39]. $\lambda'_q$ is about $1-2.2$ and $\lambda'_q$ increases with normalized plasma density in Figure 7(a). The $\lambda_q/\lambda_{q,HD}$ is in the range of $1.8-10.5$. The average values of $(\lambda_q/\lambda_{q,HD})/\lambda'_q$ for 3 plasma scenarios are 2.9, 3.6 and 4.7, respectively. The increasing $(\lambda_q/\lambda_{q,HD})/\lambda_q'$ and turbulent transport (in edge and near SOL) with higher density (Figures 7 and 2) indicate the impact of larger turbulence transport on broadening the heat flux width, especially for the high-density scenario approaching the density limit.

A major difference between $\lambda_q'$ and $\lambda_q/\lambda_{q,HD}$ may be that plasma is in H-mode with much lower turbulence level in the boundary with much stronger edge shear flow for GHD model than that for Ohmic plasma in these experiments. Anyway, it also indicates that SOL turbulent transport can play important roles in broadening the heat flux width, especially with enhanced turbulent transport in the boundary for approaching the density limit. Considering the origin of SOL turbulence and turbulent transport, we will study the impact of edge turbulence spreading on the origin of SOL turbulence and broadening of the heat flux width with increasing density.

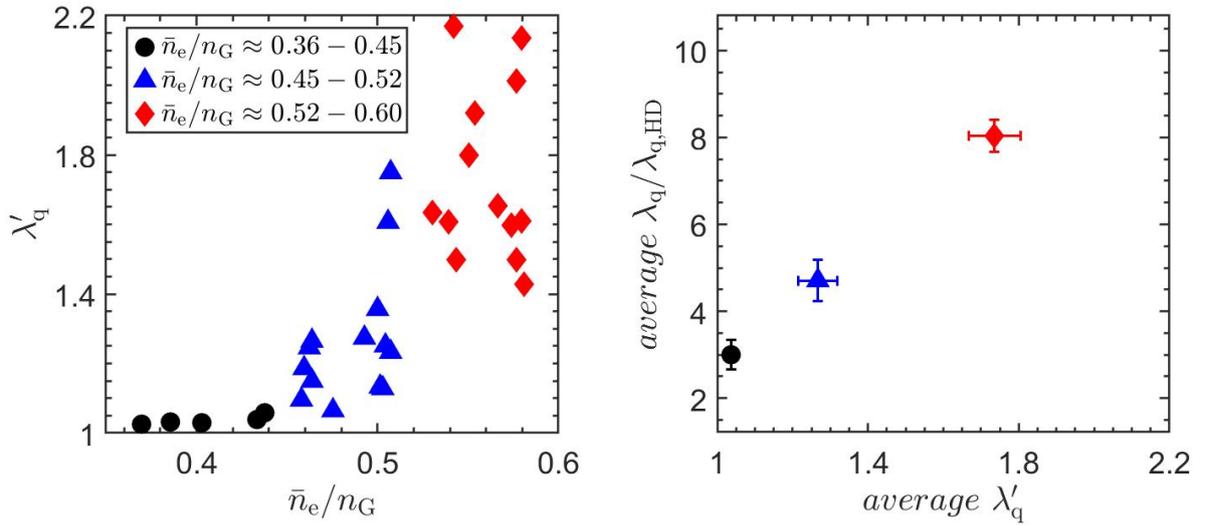

Figure 7. (a) The collisional broadening factor over the HD value $\lambda_q'$ and (b) the average ratios of experimental heat flux widths to the HD width $\lambda_q/\lambda_{q,HD}$ versus the average $\lambda_q'$ with increasing normalized density $\bar{n}_e/n_G$.

## 5  Impact of turbulence spreading on the heat flux width with increasing density

The impact of turbulence spreading on the heat flux width with increasing density is investigated in this section. Figure 8 shows that the heat flux widths increase with radial flux of turbulence internal energy near LCFS, suggesting that the heat flux width is positively correlated with edge turbulence spreading. The energy production ratio model [43] is used to explore the mechanism of turbulence spreading on broadening the heat flux width.

Figure 9(a) shows that $R_a$ is in the range of 1.6–25, which is larger than 1, indicating that edge turbulence spreading dominates the SOL turbulence. As the plasma density approaches the density limit, the average $R_a \approx$ 16 and hence $R_\text{a} \gg 1$, suggesting that edge turbulence spreading plays a much more important role than the SOL local production in the origin of the SOL turbulence in the cases in this paper. Figure 9(b) show that the heat flux widths increase with $R_\text{a}$, indicating that edge turbulence spreading dominates the SOL turbulence and hence broadens the heat flux width. Thus, edge turbulence spreading has significant impact on broadening the heat flux width, especially when plasma density approaches the density limit.

In addition, the edge $\boldsymbol{E_\text{r} \times B}$ shear influence the $R_\text{a}$. So, relations of $R_\text{a}$, edge turbulence spreading and local SOL production to the $\boldsymbol{E_\text{r} \times B}$ shearing rate are elucidated in Figure 10. $R_\text{a}$ and edge turbulence spreading decreases with the $\boldsymbol{E_\text{r} \times B}$ shearing rate while the local SOL interchange production is affected slightly by the $\boldsymbol{E_\text{r} \times B}$ shearing rate, indicating that edge turbulence spreading is more sensitive to the $\boldsymbol{E_\text{r} \times B}$ shearing rate than the local SOL production. Results in this section are consistent with findings in [43], further confirming the validity of energy production ratio model for the origin of SOL turbulence on the HL-2A tokamak and the J-TEXT tokamak.

Furthermore, edge turbulence spreading increases strongly with density approaching the density limit (Figures 4 and 9(a)), and edge $\boldsymbol{E_\text{r} \times B}$ shear flow is tightly related to edge turbulence spreading (Figures 3, 6 and 10). This indicates that for the high-density scenario, it is possible to enhance the edge $\boldsymbol{E_\text{r} \times B}$ shear flow to control the edge turbulence spreading [27], so as to balance the heat exhaust in the divertor targets and good core plasma performance.

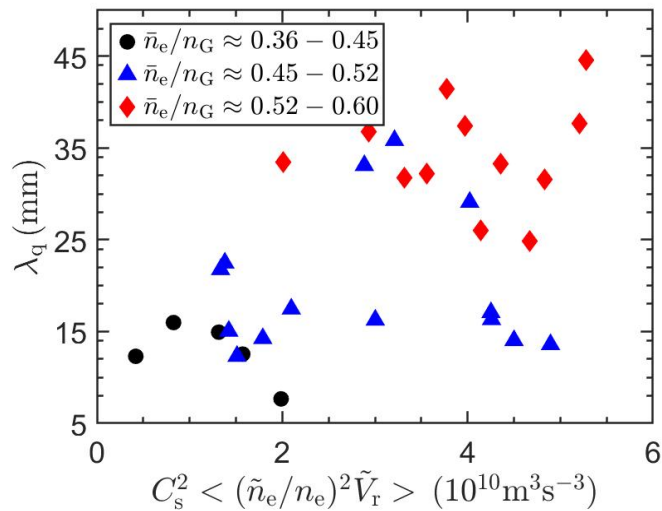

Figure 8. The relations of the heat flux widths $\lambda_q$ to radial flux of turbulence internal energy $C_s^2 \langle (\tilde{n}_e/n_e)^2 \tilde{V}_r \rangle$ near LCFS with increasing normalized density $\bar{n}_e/n_G$.

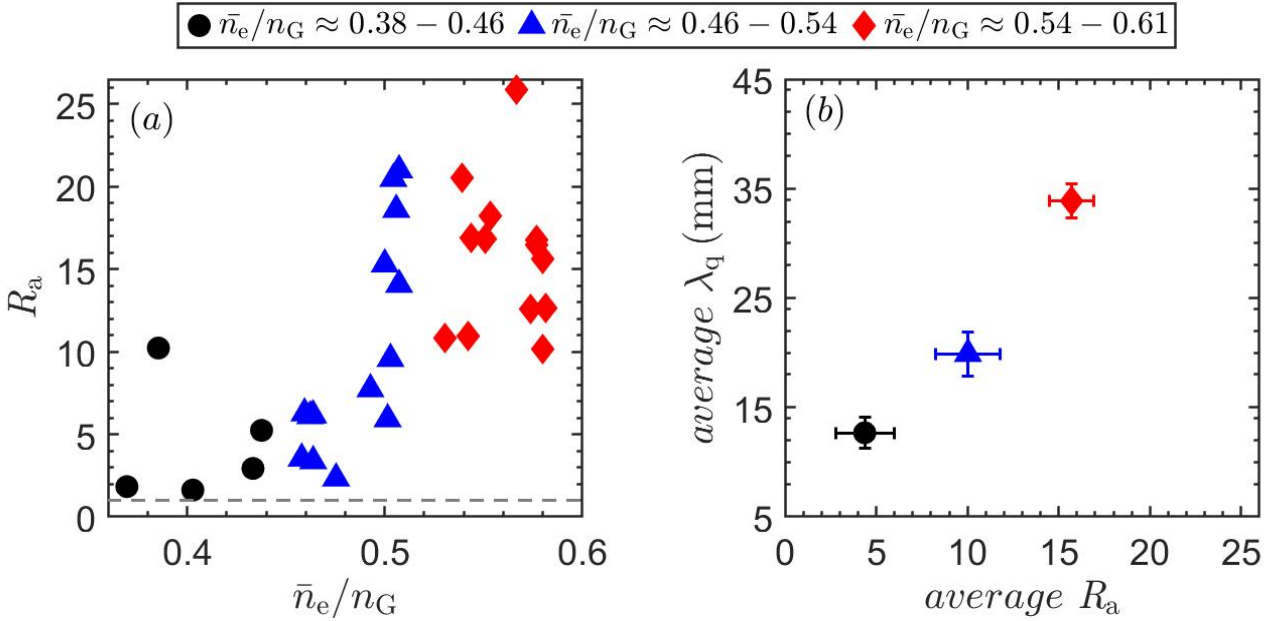

Figure 9. (a) The energy production ratio $R_a$ and (b) the relations of average heat flux widths $\lambda_q$ to average energy production ratio $R_a$ with increasing normalized density $\bar{n}_e/n_G$.

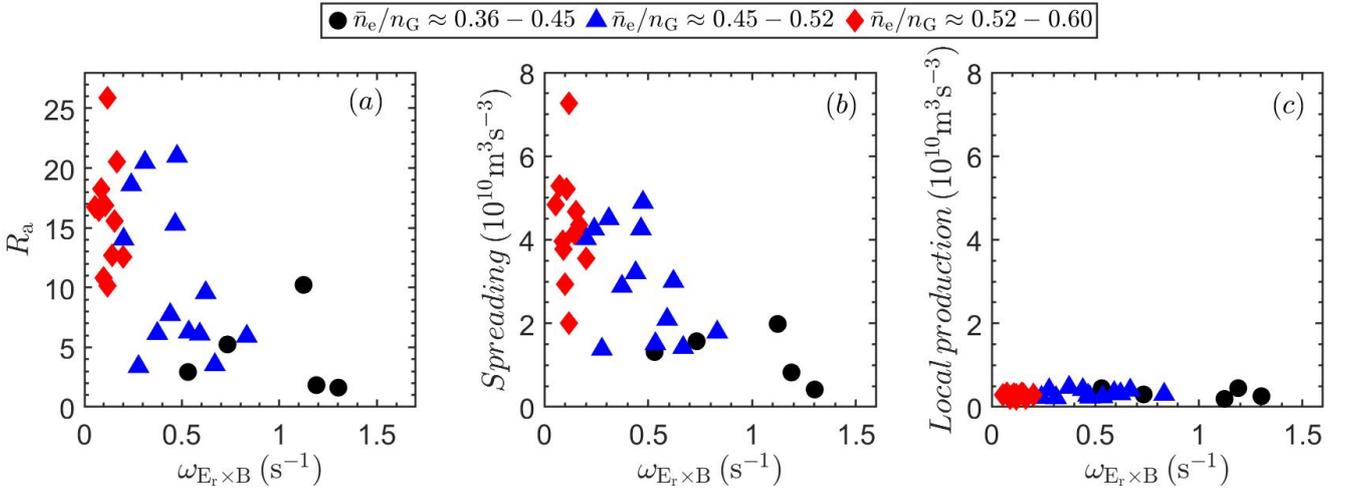

Figure 10. The relations of (a) energy production ratio $R_a$, (b) turbulence spreading at LCFS and (c) local SOL production to the $E_r \times B$ shearing rate $\omega_{E_r \times B}$ with increasing normalized density $\bar{n}_e/n_G$.

## 6  Edge turbulence spreading by blobs and their roles in the heat flux widths

The impact of blob transport on the heat flux width is investigated with increasing density in this section. Blob dynamics is an essential part of turbulent transport at plasma boundary [33–36], especially when plasma density approaches the density limit [26, 28]. In this paper, a "blob" is identified when the fluctuation amplitude in ion saturation current exceeds a threshold value (2.5 times its standard deviation in this paper) [43]. Blob behaviour is usually described by skewness, turbulent particle flux and turbulence spreading rate [33]. In addition, the ratio of blob-induced turbulent particle flux to the total turbulent particle flux ($\Gamma_{\text{blob}}/\Gamma_{\text{total}}$) is used to quantify the contribution of blob particle transport to the total turbulent particle transport. Here $\Gamma_{total} = \int \Gamma(t)\,dt$ with $\Gamma(t) = \widetilde{V}_r \widetilde{n}_e$, and $\Gamma_{\text{blob}} = \sum_i \Gamma(t_i - \tau/2 < t < t_i + \tau/2)$ (here i is the number of blobs; $t_i$ is the peak time point of each blob; $\tau$ is the lifetime of a blob estimated by the conditional average method [59–60]). Furthermore, the ratio of blob-induced turbulence spreading to the total spreading ($Sp_{\text{blob}}/Sp_{\text{total}}$) is also investigated to quantify the effect of turbulence spreading. $Sp_{\text{blob}}/Sp_{\text{total}}$ is calculated as $C_s^2 (\widetilde{n}_e/n_e)^2 \widetilde{V}_r|_{\text{blob}} / C_s^2 (\widetilde{n}_e/n_e)^2 \widetilde{V}_r|_{\text{total}}$, with the same method to estimate $\Gamma_{\text{blob}}/\Gamma_{\text{total}}$.

In Figure 11, edge skewness, conditional averaged blob particle flux and turbulence spreading rate increase with $\bar{n}_e/n_G$. In Figures 12(a) and 12(b), $\Gamma_{\text{blob}}/\Gamma_{\text{total}}$ is 0.2–0.4 while $Sp_{\text{blob}}/Sp_{\text{total}}$ is 0.5–0.9. Especially, the average $Sp_{blob}/Sp_{total}$ is as much as 0.81 for the high-density scenario, emphasizing the dominant role of spreading induced by blob in the total edge spreading. $\Gamma_{\text{blob}}/\Gamma_{\text{total}}$ and $Sp_{\text{blob}}/Sp_{\text{total}}$ increase with density, indicate that blob transport becomes stronger with increasing density, consistent with results in [26, 28]. This also indicates that spreading is a more important parameter than turbulent particle flux to represent blob transport. In Figures 12(c) and 12(d), $\lambda_q$ increase with $\Gamma_{\text{blob}}/\Gamma_{\text{total}}$ and $Sp_{\text{blob}}/Sp_{\text{total}}$. This indicates that blob particle transport, especially blob spreading plays important roles in broadening the heat flux width.

Blobs radial scale ($\delta_{\text{r,blob}}$) with increasing density are examined in figure 13. The blob radial scale is estimated by blob lifetime times its radial velocity using conditional average method [?]. $\delta_{\text{r,blob}}$ is 2–16 mm in the edge, which is about 1/3 of the heat flux width. $\delta_{\text{r,blob}}$ increases with normalized density. Moreover, the heat flux widths correlate positively with blob radial scales. Figures 12–13 indicate that, larger blob radial scale carries larger spreading into the SOL, thus dominating the SOL turbulence. Consequently, heat flux width is broadened when plasma density approaches density limit.

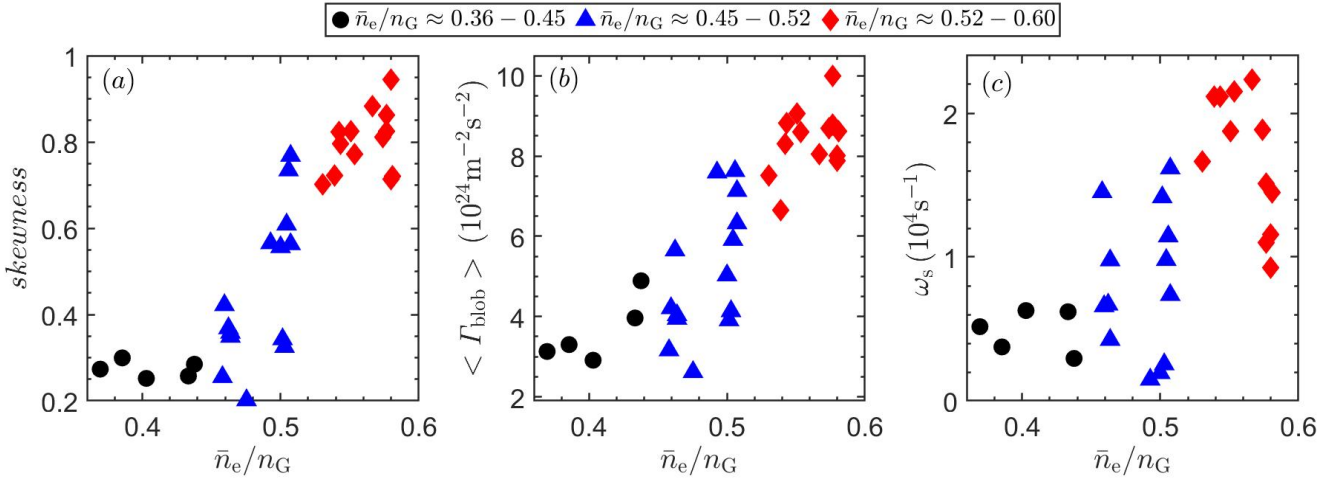

Figure 11. (a) The skewness, (b) blob-induced turbulent particle flux by conditional average method $\Gamma_{\text{blob}}$ and (c) the turbulence spreading rate $\omega_s$ with increasing normalized density $\bar{n}_e/n_G$.

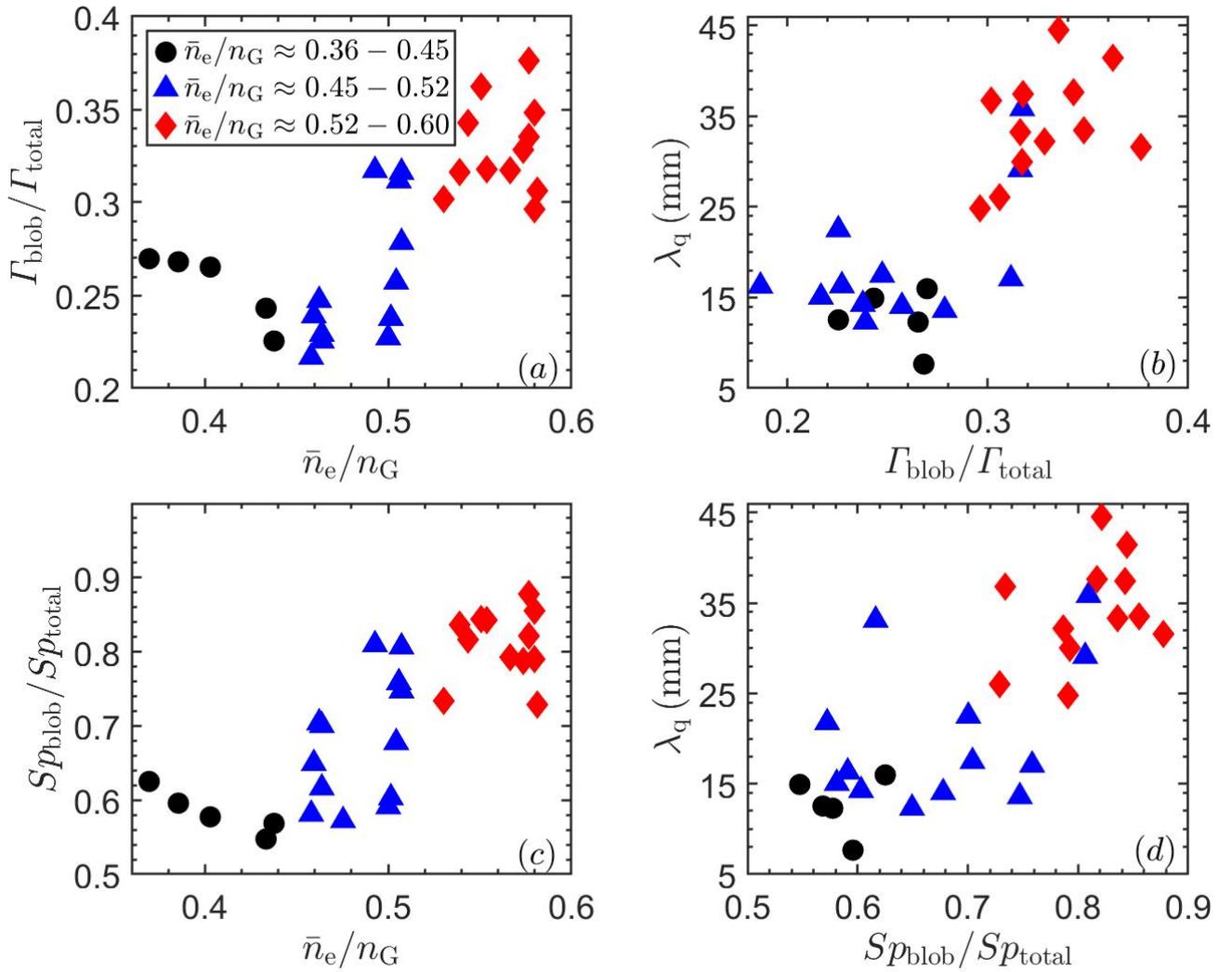

Figure 12. (a) $\Gamma_{\text{blob}}/\Gamma_{\text{total}}$ and (c) $Sp_{\text{blob}}/Sp_{\text{total}}$ with increasing normalized density $\bar{n}_e/n_G$. The heat flux width $\lambda_q$. versus (c) $\Gamma_{\text{blob}}/\Gamma_{\text{total}}$ and (d) $Sp_{\text{blob}}/Sp_{\text{total}}$.

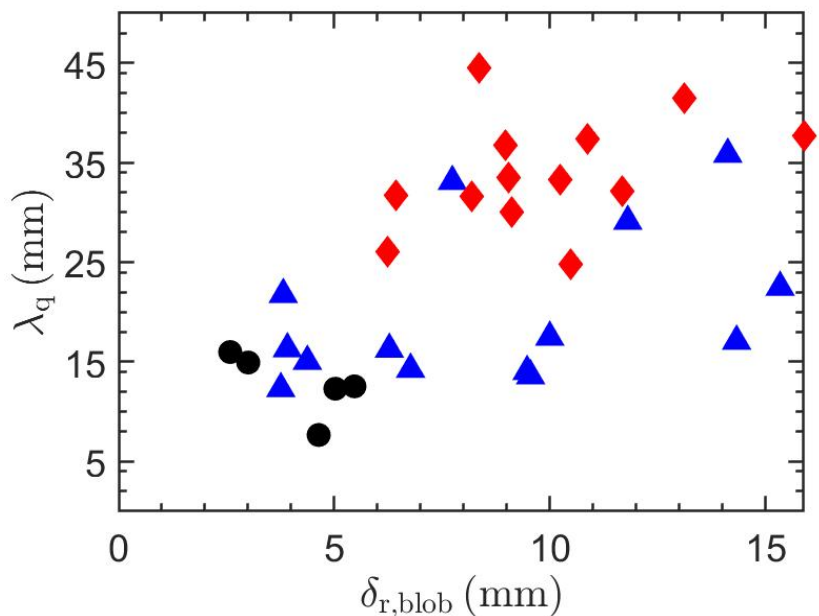

Figure 13. The relation of the heat flux width $\lambda_q$ to the blob radial scale $\delta_{r,\text{blob}}$.

## 7  Conclusion and further study

This paper investigates the impact of edge turbulence spreading on broadening the heat flux width in Ohmic-plasma approaching the density limit on the J-TEXT tokamak. The principal results are as follows.

i) When plasma approaches the density limit, the edge $E_r \times B$ shear flow collapses, and edge turbulent transport and turbulence spreading are greatly enhanced. Plasma converts from the adiabatic regime ($\alpha > 1$) to hydrodynamic regime ($\alpha < 1$).

ii) The heat flux width increases with the edge fluctuation, turbulence auto-correlation time, turbulent particle transport, while decreases with the edge $E_r \times B$ shearing rate. The heat flux widths are broadened to 3–7 times of the widths predicted from GHD model.

iii) The heat flux width increases with radial flux of turbulence internal energy, namely turbulence spreading. The energy production ratio is much larger than 1 for high-density scenario, suggesting edge turbulence spreading dominates the origin of SOL turbulence. The heat flux width increases with the energy production ratio, indicating that larger edge turbulence spreading broadens the heat flux width. In addition, edge turbulence spreading is more sensitive to the edge $E_r \times B$ shear flow than the local SOL interchange production.

iv) The impact of blob-induced transport on the heat flux width is investigated with increasing normalized density. The edge skewness, conditional average particle flux, turbulence spreading rate of blobs increase with higher normalized density. Blob-induced spreading ($C_s^2(\tilde{n}_e/n_e)^2 \tilde{V}_r|_{\text{blob}}$) is about 81% of total turbulent spreading ($C_s^2(\tilde{n}_e/n_e)^2 \tilde{V}_r|_{\text{total}}$), thus playing important roles in the SOL turbulence. Blob with larger radial scales carry stronger edge spreading into the SOL, dominating SOL turbulence and consequently broadening the heat flux width as plasma density approaches the density limit.

Over all, edge turbulence spreading dominates the SOL turbulence and plays a significant role in broadening the heat flux width when plasma approaches the density limit. Edge $\mathbf{E_r} \times \mathbf{B}$ shear flow is tightly related to edge turbulence spreading. This paper indicates that, regulating the edge $\mathbf{E_r} \times \mathbf{B}$ shear flow to control edge turbulence spreading and hence the heat flux width, may be a key factor to balance core confinement with heat exhaust in the divertor targets for high-density high-performance plasma.

Several directions for future work are indicated. First, since the edge turbulence spreading has strong impact on broadening the heat flux width, the relation of the heat flux width to the edge turbulence spreading $\lambda_q = \lambda(\Gamma_E, parameters)$ should be deduced. Second, the impact of blob propagation from the edge to the near SOL on the broadening of the heat flux width requires further study.

## Acknowledgements

This work is supported by National Key Research and Development Program of China (Nos. 2022YFE03020001), National Natural Science Foundation of China (Nos. 12305235, 12435015, 12305238, 12305237 and 12375210), the Sichuan Youth Science and Technology Innovation Team Project under Grant No. 2022JDRC0014) and Hubei International Science and Technology Cooperation Projects (No. 2022EHB003). US DoE Grant DE-FG02-04ER54738 and the Sci DAC ABOUND Project, scw1832 are also acknowledged.

## Appendix A

This appendix aims to compare the heat flux widths from experimental measurements and scaling laws in reference [6-8, 15]. We use experimental data in limited Ohmic-plasma on J-TEXT to calculate the heat flux widths from these scaling laws. The heat flux widths from the 5 scaling laws are 12.5 mm, 63.6 mm, 47.3 mm, 21.6 mm and 28.5 mm respectively, shown in table 2. The difference in heat flux width from 5 scaling laws may relate to different types of SOL turbulence and requires further study. The experimental heat flux widths range from 8 mm to 45 mm, with an average value 24.5 mm. Moreover, the scaling laws in table 2 do not depend on the plasma density, indicating that these scaling laws may not apply to the limited Ohmic-plasma on J-TEXT, especially in the high-density regime.

Table 2. The heat flux widths for different scaling laws and experiments in J-TEXT limited Ohmic-plasma.

| Heat flux widths | $\lambda_q$ (mm) |
| --- | --- |
| Scaling laws 1: $\lambda_q(\text{m}) = 10 \times \left(P_{tot}/V(\text{Wm}^{-3})\right)^{-0.38} (a/R/\kappa)^{1.3}$  [8] | 12.5 |
| Scaling laws 2: $\lambda_q(\text{cm}) = 3.1 \times (P_{tot}(\text{MW}))^{-0.46}$  [6] | 63.6 |
| Scaling laws 3: $\lambda_q(\text{cm}) = 2.1 \times (P_{tot}(\text{MW}))^{-0.52}$  [6] | 47.3 |
| Scaling laws 4: $\lambda_q(\text{m}) = 8.7 \times R^{0.73} q^{0.76} B_t^{-0.29}$  [7] | 21.6 |
| Scaling laws 5: $\lambda_q(\text{mm}) = 5 \times q^{0.68} B^{-0.66} R^{0.34} (A/Z)^{0.33} \left(P_{sep}\right)^{0.09} a^{-0.09}$  [15] | 28.5 |
| Exp. Heat flux widths | 8 - 45 |